\documentclass{article}
\usepackage{amsmath}
\usepackage{graphicx}
\usepackage{natbib}
\usepackage{url} 
\usepackage{graphicx}
\usepackage{amsmath}
\usepackage{natbib}
\usepackage{amscd,amssymb,amsfonts,verbatim}
\usepackage{mathrsfs}
\usepackage{bbm}
\usepackage{listings}
\usepackage{epsfig}
\usepackage{enumitem}

\usepackage{url}
\usepackage{booktabs,makecell,ltablex}
\usepackage{multirow}

\usepackage{tikz}
\usetikzlibrary{patterns}
\usetikzlibrary{trees}
\usepackage{pgfplots}
\usetikzlibrary{svg.path}
\usepgfplotslibrary{fillbetween}
\usetikzlibrary{shapes.misc}
\usetikzlibrary{shapes.geometric}
\usetikzlibrary{arrows}
\pgfplotsset{compat=1.16}
\pgfplotsset{soldot/.style={color=blue,only marks,mark=*}}
\pgfplotsset{holdot/.style={color=blue,fill=white,only marks,mark=*}}
\newcommand\BibTeX{{\rmfamily B\kern-.05em \textsc{i\kern-.025em b}\kern-.08em
		T\kern-.1667em\lower.7ex\hbox{E}\kern-.125emX}}

\def\blambda{\boldsymbol{\lambda}}
\def\bTheta{\boldsymbol{\Theta}}

\def\bQ{\mathbf{Q}}

\def\bB{\mathbf{B}}

\newcommand{\cK}{\mathcal{K}}
\newcommand{\Nall}{N_{[0,\tau]}}
\newcommand{\Xall}{X_{[0,\tau]}}

\newcommand{\xall}{x_{[0,\tau]}}
\newcommand{\Xalln}{X^n_{[0,\tau^n]}}
\newcommand{\xalln}{x^n_{[0,\tau^n]}}

\graphicspath{{Figs/}}

\makeatother

\begin{document}

\def\spacingset#1{\renewcommand{\baselinestretch}%
{#1}\small\normalsize} \spacingset{1}


  \title{\bf Bayesian inference for the Markov-modulated Poisson process with an outcome process}
  \author{Yu Luo\thanks{
  		Department of Mathematics, King's College London, United Kingdom} ,  \hspace{.2cm}
  	Chris Sherlock\thanks{
  		Department of Mathematics and Statistics, Lancaster University, United Kingdom} \hspace{.2cm}
  }
 \date{ }
  \maketitle

\bigskip
\begin{abstract}
In medical research, understanding changes in outcome measurements is crucial for inferring shifts in health conditions. However, traditional methods often struggle with large, irregularly longitudinal data and fail to account for the tendency of individuals in poorer health to interact more frequently with the healthcare system. Additionally, clinical data can lack information on terminating events like death. To address these challenges, we start from the continuous- time hidden Markov model which models observed data as outcomes influenced by latent health states. Our extension incorporates a point process to account for the impact of health states on observation timings and includes a ``death" state to model unobserved terminating events through a Poisson process, where transition rates depend on the latent health state. This approach captures both the severity of the disease and the timing of healthcare interactions. We present an exact Gibbs sampler procedure that alternates between sampling the latent health state paths and the model parameters. By including the ``death" state, we mitigate biases in parameter estimation that would arise from solely modelling ``live" health states. Simulation studies demonstrate that the proposed Gibbs sampler performs effectively. We apply our method to Canadian healthcare data, offering valuable insights for healthcare management.
\end{abstract}

\noindent%
\vfill

\newpage
\spacingset{1.5}

\section{Introduction}\label{sec1}
In medical research, observed changes in outcome measurement are used to infer changes in the underlying condition of the patient. Data must be collected over time (longitudinally) on a patient to characterize the effect of risk factors and to evaluate the effectiveness of treatments. For instance, understanding the progression of chronic diseases, such as chronic obstructive pulmonary disease (COPD), is important to inform early diagnosis, personalized care and health system management. With no existing cure, the severity of COPD increases over time, producing more frequent exacerbations and increasing use of healthcare \citep{burge2003copd}. This worsening disease trajectory defines COPD and many other chronic illnesses. Despite this natural history of worsening disease, health system indicators currently describe COPD from a cross-sectional perspective. These traditional measures of service volume, such as hospitalization rates, neglect the temporal disease progression within individuals that is central to managing COPD. Data from clinical and administrative systems have the potential to advance this understanding, but traditional methods for modelling disease progression are not well-suited to analyzing high-dimensional data collected at irregular intervals. Many cases of severe illness can be traced back to a lack of appropriate and ongoing follow-up health care. To help understand the long-term causes of severe and preventable illness, we need to understand not only how many patients have severe illness, but also what their ``care paths" or ``trajectories" were before they ended up with severe illness. Private and public health care systems now have the ability to maintain longitudinal digital health records. These data from sources, such as electronic health records, health systems and data from mobile health applications, allow the study of subject-specific trajectories, which can inform clinical and public health decision-making. However, data are typically recorded only when a subject interacts with a provider, resulting in irregularly-spaced longitudinal observations, and the patterns of clinical interactions vary from subject to subject. Although the data allow dynamic monitoring of the underlying disease progression that governs the data recorded in the health system, this disease progression cannot be observed directly, and so inferential methods are needed to ascertain the latent progression. The statistical analysis of such data, especially when the cohort is large and observations are frequent, is complex.

A multi-state model captures the status of an individual longitudinally as a time-discretization of a continuous-time Markov process, and in many settings the common assumption that the health status of a patient falls into one of a finite number of categories (usually ordered by severity) is a plausible one. Inference for continuous-time Markov processes observed discretely and either precisely or partially has been well studied \citep[see, for example,][]{bladt2005statistical, pfeuffer2017ctmcd}. However, health system records typically do not include direct information about underlying health status; instead, it is inferred indirectly through a proxy measure. Such a process can be modelled as a continuous-time hidden Markov model \cite[CTHMM, e.g.][]{CapMouRyd2005}, which has been studied recently in medical applications such as disease progression  \citep{luocthmm2018a} and cancer screening \citep{lange2015joint,klausch2023bayesian}. While a CTHMM can address the issue of non-equidistant longitudinal observations, in medical research,  it is possible that the timing of healthcare utilization itself is informative in the assessment of health status progression. For example, patients who are in more severe health states typically interact with the healthcare \citep[e.g.][]{cook2018multistate}. However, if the intensity of the Poisson process depends on the state of a hidden Markov process, it is often referred to as a Markov-modulated Poisson process \cite[e.g.][]{ryden1996algorithm,fearnhead2006exact}.  There have been studies modelling this observation time process using frequentist approaches,  such as \cite{lange2015joint} and \cite{mews2023markov}. 

In addition, data from the claim and health care system provide no information for terminating events, such as death. However, appropriate parameter inference will only follow if we distinguish a lack of interaction with the healthcare system due to, for example, mild illness from a lack of interaction due to the subject being dead. \cite{Nevala2024} included the death data from a different data system to model colorectal cancer progression. However, to the best of our knowledge, there is no existing method to account for the terminating event, such as death, when it is not available in a dataset and event times are informative. In this paper, we develop a Bayesian inference procedure that takes into account the observation process by modelling the observed data as an outcome process and the time of observations as a point process jointly alongside the CTHMM, given the underlying latent health state. Moreover, we also consider the death time as missing data, and model the death as an informative censoring.  This extension allows us to model disease severity and death not only from the types of care received but also from the times between different observed events and their respective frequency. It also provides insights on how long patients tend to survive in each state. 

The paper is organized as follows. Section \ref{MMPP} demonstrate the model structure, followed by the likelihood construction in Section \ref{lik} and the forward-backward method to calculate the probability of each state at the observation time in Section \ref{fbal}. Section \ref{Gibbs} layouts the Gibbs sampler to draw inference for the model, and some empirical studies, including simulation and real data analysis, are in Section \ref{emprical}. The paper concludes in Section \ref{dis} with a summary of findings and discusses areas for future research.

\section{A Markov-modulated Poisson process with an outcome process}
\label{MMPP}
In this section, we describe the continuous-time hidden Markov model with the observation time following a Poisson process. In the sequel, for any collection of scalars, $a_1,\dots,a_T$ or $a^1,\dots,a^T$, we denote the concatenations by $a_{1:T}$ and $a^{1:T}$, respectively. Thus, for a particular individual we observe the data $o_{1:T}:=\{o_1,\ldots,o_T\}$ at  time points $\tau_{1:T}:=\{\tau_1,\dots,\tau_T\}$. We assume that the data arise as a consequence of a latent continuous-time Markov chain (CTMC) $\{X_s\}_{s\in[0,\tau]}$ taking values on the finite integer set $\cK:=\{1,2,\ldots,K\}$. We denote the generator of the latent process by $\bQ$ and the initial distribution by $\nu$. At  observation time $\tau_t$ ($t=1,\dots,T$) the latent process is in state $X_{\tau_t}$ and the observation is $o_t$.

Conditional on the latent process being in state $k\in \cK$, at time $\tau_t$, the outcome observation $o_t$ is drawn from a distribution with a probability mass or density function $f_t(o_t|X_{\tau_t}=k)\equiv f_k(o_t|z_t,\beta_k)$, where $z_t\in \mathbb{R}^D$ is a vector of covariates (including an intercept) and $\beta$ is a $D$-vector of covariate effects. In all of our simulations, including the real-data example, the prior for $\beta_k$, is chosen to be conditionally conjugate to $f_k$ so that it is straightforward to sample from the conditional posterior for each $\beta_k$ given $X_{\tau_1},\dots,X_{\tau_T}$. It is possible to extend inference to the non-conjugate case using the Metropolis--Hastings algorithm; see the supplementary material for a brief discussion. Define the matrix $\bB = (\beta_{d,k})$ for $d=1,\ldots,D$ and $k =1,\ldots,K$ as the regression coefficient matrix.   

In addition, we allow the observation process itself (that is, the times that the observations are made) to be informative about the latent trajectory. Specifically, let $N_s$ be the number of observations (events of the Poisson process) on $[0,s]$, $N_s=\sum_{t=1}^T 1(\tau_t\le s)$, so that $N_0=0$ and $N_{\tau_t}=t$.  We assume that $N_s$ follows a Poisson process whose intensity is $\lambda_k$ when $X_s=k$. The joint process $(N_s,X_s)$ is a Markov modulated Poisson process (MMPP) and inference approaches have been extensively studied in, for example,  \cite{ryden1996algorithm,fearnhead2006exact}. To simplify notation we write $\Xall$ for $\{X_s\}_{s\in [0,\tau]}$ and $\Nall$ for $\{N_s\}_{s\in [0,\tau]}$. 

Conditional on $X_s$, the observations $O_{1:T}$ are assumed to be independent of each other and of $N_s$; also $N_s$ is assumed to evolve independently of $X_s$ except that its intensity is $\lambda_{X_s}$.
Therefore, the observation time $\tau_t$ is generated from the MMPP $(N_s,X_s)$ and the outcome $O_t$ is then generated from $f_{X_{\tau_t}}(o_t|z_t,\beta_{X_{\tau_t}})$. Figure \ref{tikzfig:data} provides a schematic of the presumed data generating structure for one subject. Conditional on $\nu$, $\bQ$, $\blambda:=(\lambda_1,\dots,\lambda_k)$ and $\bB$, data for each subject are assumed to be generated independently of the data for all other subjects. Ideally we observe the process over some time interval $[0,\tau]$ for some $\tau>\tau_T$ and with $\tau_1>0$. However, for each patient, our data start with a diagnosis of COPD, so $\tau_1=0$. 

\tikzset{filled/.style={fill=circle area, draw=circle edge, thick},
	outline/.style={draw=circle edge, thick}}
\tikzset{cross/.style={cross out, draw=black, minimum size=2*(#1-\pgflinewidth), inner sep=0pt, outer sep=0pt},
	cross/.default={1pt}}
\tikzset{vertex/.style = {shape=circle,draw,minimum size=1.5em}}
\tikzset{edge/.style = {->,> = latex'}}
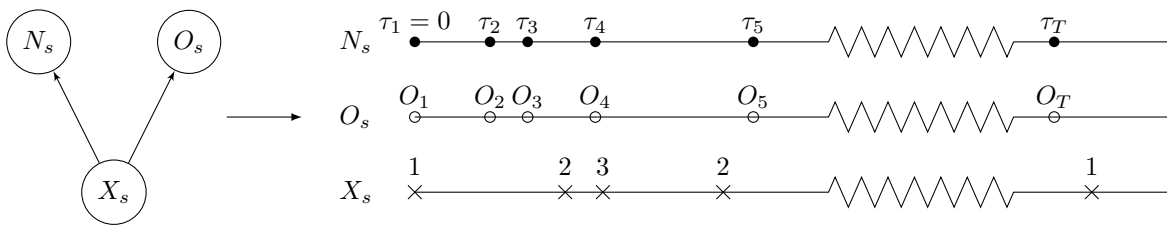
\begin{figure}[ht]
	\centering
	\begin{tikzpicture}
		\tikzset{edge/.style = {->,> = latex'}}
		\draw[decorate,
		decoration={zigzag,
			pre length=5.5cm,
			post length=2cm,
			amplitude=2mm
		}] (0,0) -- (10,0);
		\draw[decorate,
		decoration={zigzag,
			pre length=5.5cm,
			post length=2cm,
			amplitude=2mm
		}] (0,1) -- (10,1);
		\draw[decorate,
		decoration={zigzag,
			pre length=5.5cm,
			post length=2cm,
			amplitude=2mm
		}] (0,2) -- (10,2);
		
		\draw (0,0)  node[cross=3pt]{};
		\draw (0,0.1)  node[above] {1};
		\draw (2,0)  node[cross=3pt]{};
		\draw (2,0.1)  node[above] {2};
		\draw (2.5,0)  node[cross=3pt]{};
		\draw (2.5,0.1)  node[above] {3};
		\draw (4.1,0)  node[cross=3pt]{};
		\draw (4.1,0.1)  node[above] {2};
		\draw (9,0)  node[cross=3pt]{};
		\draw (9,0.1)  node[above] {1};

		\fill (0,2)  circle[radius=2pt] node[above] {$\tau_1=0$};
		\fill (1,2)  circle[radius=2pt] node[above] {$\tau_2$};
		\fill (1.5,2)  circle[radius=2pt] node[above] {$\tau_3$};
		\fill (2.4,2)  circle[radius=2pt] node[above] {$\tau_4$};
		\fill (4.5,2)  circle[radius=2pt] node[above] {$\tau_5$};
		\fill (8.5,2)  circle[radius=2pt] node[above] {$\tau_T$};

		\draw (0,1)  circle[radius=2pt] node[above] {$O_1$};
		\draw (1,1)  circle[radius=2pt] node[above] {$O_2$};
		\draw (1.5,1)  circle[radius=2pt] node[above] {$O_3$};
		\draw (2.4,1)  circle[radius=2pt] node[above] {$O_4$};
		\draw (4.5,1)  circle[radius=2pt] node[above] {$O_5$};
		\draw (8.5,1)  circle[radius=2pt] node[above] {$O_T$};
		
		\node[text width=1cm] at (-0.5,2) {$N_s$};
		\node[text width=1cm] at (-0.5,1) {$O_s$};
		\node[text width=1cm] at (-0.5,0) {$X_s$};

		\node[vertex] (n1) at (-5,2) {$N_s$};
		\node[vertex] (o1) at (-3,2) {$O_s$};
		\node[vertex] (x1) at (-4,0) {$X_s$};
		\draw[edge] (x1) to (o1);
		\draw[edge] (x1) to (n1);
		
		\draw[>=latex, ->] (-2.5,1) -- (-1.5,1);	
	\end{tikzpicture}
	\caption{Schematic of the presumed data generating mechanism for the MMPP with an outcome process $O_{\tau_t}$.  $X_s$ represents the process underlying the observed data, with observation (event) time points denoted $\tau_t, t=1,\ldots,T$ for a state-dependent Poisson process $N_s$; $X_s$ represents the hidden Markov process.\label{tikzfig:data}}  
\end{figure}

\subsection{Likelihood for the MMPP with an outcome process}
\label{lik}
We are interested in inference for the data generated from the the MMPP with an outcome process.  We now construct the complete-data likelihood for one subject over the time interval $\left[0,\tau \right]$. Since subjects are independent, the likelihood for all subjects is the product of the individual likelihoods.

The likelihood function for $\bQ$ and $\nu$ is \citep{bladt2005statistical}
\begin{equation*}
	\mathcal{L}(\bQ,\nu|\xall) 
	=
	f(\xall|\bQ,\nu) = \nu_{x_0} \prod\limits_{l = 1}^K {\prod\limits_{m \ne l} {{q_{l,m}}^{{N_{l,m}}}\exp \left( { - {q_{l,m}}{R_{l}}} \right)} },
\end{equation*}
where $N_{l,m}$ is the number of transitions from state $l$ to state $m$ in the time interval $\left[0,\tau\right]$ and $R_{l}$ is the total time that the process has spent in state $l$ in $\left[0,\tau\right]$. The values of $q_{l,m}$ are the elements in $\bQ$. In addition, the likelihood function for $\blambda$ is 
\begin{equation*}
	\mathcal{L}(\blambda; \xall,\tau_{1:T})=f(\tau_{1:T}|\xall,\blambda) = \prod\limits_{i = 1}^K {{\lambda_{i}}^{{N_{i}}}\exp \left(  - {\lambda_{i}}{R_{i}} \right)} 
\end{equation*}
where $N_{i}$ is the number of events at state $i$. Note that the quantities $N_{l,m}$, $R_{l}$ and   $N_{i}$ all are unobserved, but they can be computed when the underlying process $\Xall$ is known.

The complete likelihood, derived from $o_{1:T}$, $\tau_{1:T}$ and $\Xall$ can be factorized:
\[
\mathcal{L}(\bB,\blambda,\bQ,\nu)
\equiv
f(o_{1:T},\tau_{1:T},\xall|\bB,\blambda,\bQ,\nu)
=
f(o_{1:T}|\bB,\xall)
f(\tau_{1:T}|\xall,\blambda)f(\xall|\nu,\bQ).
\]
Thus
\begin{equation}
	\label{eqlik}
	\mathcal{L}(\bB, \blambda,Q,\nu)
	= \prod\limits_{t = 1}^{T} {f_{X_{\tau_t}}\left( {o_{t}\left| {z_t,\beta_{x_{\tau_{t}}}} \right.} \right)} \times\nu_{x_{0}}  \prod\limits_{l = 1}^K {\prod\limits_{m \ne l} {{q_{l,m}}^{{N_{l,m}}}\exp \left( { - {q_{l,m}}{R_{l}}} \right)} } \prod\limits_{i = 1}^K {{\lambda_{i}}^{{N_{i}}}\exp \left(  - {\lambda_{i}}{R_{i}} \right)} . 
\end{equation}

If independent Gamma-distributed priors are assigned to all of the components $\lambda_i$ and $q_{l,m}$ (see Section \ref{sec.priors}) then, conditional on the  complete chains for all individuals, the posteriors for these parameters also have independent Gamma distributions. This permits the use of the Gibbs steps described in Section \ref{sec.GS}. An alternative partial imputation, which necessitates Metropolis--Hastings updates for these parameters, is discussed in the Supplementary Material.

\subsection{Data-induced window restriction}
\label{sec.windowRestrict}
Our data are of interactions between patients with COPD and the health service during a 17-year window between 1st January 1998 and 31st December 2014. If a patient was diagnosed with COPD within the window then the first observation for that patient is the time of diagnosis. The patient may have interacted with the health service within the 17-year window and before they were diagnosed with COPD, but we do not know about these interactions. Therefore, for any given patient, we set $\tau_1=0$; we start the clock at the first interaction. Since we know there will always be an interaction at time 0, this observation does not contribute to the count $N_k(\tau)$ in \eqref{eqlik}. 
Of necessity, therefore, the initial distribution $\nu$ represents an average of the distribution of the state when an individual is first diagnosed with COPD and, for those already had with COPD at the start of 1998, the distribution of the state at their first interaction within the window.

For each patient, we do not have the time of death or, even, an indicator of whether or not death occurred before the end of 2014. If the last interaction with the health service was just before the end of the window then it seems likely the patient was still alive at the end of the window; however, if the last interaction was well before the end of the window and there had been a high frequency of interactions before this then it is likely that the patient died or moved away from the province shortly after the final interaction. {In Section \ref{emprical}, for the data analysis and one of the simulation studies, we circumvent this issue by introducing an additional state for the hidden chain, corresponding to death (or moving away). This is an absorbing state and there are no observations whilst in this state.}

Figure \ref{extrajectory} shows three example  patients' health trajectories. The observation window is between 0 and $\tau$, and assume that all patients have an observation at time 0. Patient 1 died right before $\tau$ and we have the record for the last observation but do not have information for the death. Patient 2, on the other hand, has one additional interaction after the observation window and died afterwards. For these patients, the death happens close to the observation window. Modelling death does not have much impact on estimates of $\bQ$ or $\blambda$. However, patient 3 has an early death, and the trajectory for observed data is only half of the observation window.  For these patients who die early, in particular, we need to model death instead of, for example, assuming the patient is alive until the end of the window.

\tikzset{filled/.style={fill=circle area, draw=circle edge, thick},
	outline/.style={draw=circle edge, thick}}
\tikzset{cross/.style={cross out, draw=black, minimum size=2*(#1-\pgflinewidth), inner sep=0pt, outer sep=0pt},
	cross/.default={1pt}}
\tikzset{vertex/.style = {shape=circle,draw,minimum size=1.5em}}
\tikzset{edge/.style = {->,> = latex'}}
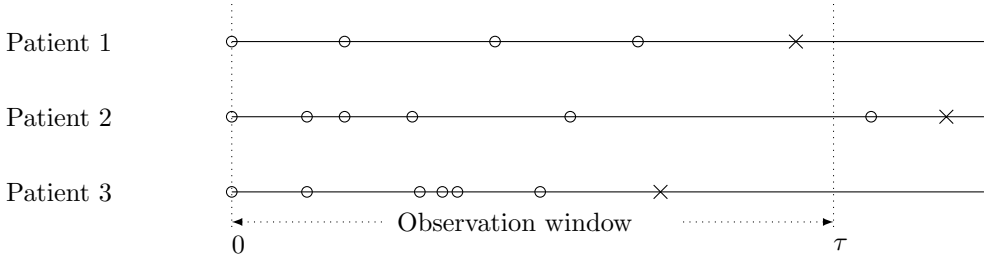
\begin{figure}[ht]
	\centering
	\begin{tikzpicture}
		\tikzset{edge/.style = {->,> = latex'}}
		\draw[decorate] (0,0) -- (10,0);
		\draw[decorate] (0,1) -- (10,1);
		\draw[decorate] (0,2) -- (10,2);
		
		\draw (0,0)  circle[radius=2pt];
		
		\draw (1,0)  circle[radius=2pt];
		\draw (2.5,0)  circle[radius=2pt];
		\draw (2.8,0)  circle[radius=2pt];
		\draw (3,0)  circle[radius=2pt];
		\draw (4.1,0)  circle[radius=2pt];
		\draw (5.7,0)  node[cross=3pt]{};

		\draw (0,2)  circle[radius=2pt] ;
		\draw (1.5,2)  circle[radius=2pt] ;
		\draw (3.5,2)  circle[radius=2pt] ;
		\draw (5.4,2)  circle[radius=2pt];
		\draw (7.5,2)  node[cross=3pt]{};

		\draw (0,1)  circle[radius=2pt] ;
		\draw (1,1)  circle[radius=2pt] ;
		\draw (1.5,1)  circle[radius=2pt] ;
		\draw (2.4,1)  circle[radius=2pt] ;
		\draw (4.5,1)  circle[radius=2pt] ;
		\draw (8.5,1)  circle[radius=2pt] ;
		\draw (9.5,1)  node[cross=3pt]{};
		\node[text width=2cm] at (-2,2) {Patient 1};
		\node[text width=2cm] at (-2,1) {Patient 2};
		\node[text width=2cm] at (-2,0) {Patient 3};
		
		\draw[dotted] (8,2.5) -- (8,-0.5);	
		\node[text width=1cm] at (8.5,-0.7) {$\tau$};
		\draw[dotted] (0,2.5) -- (0,-0.5);	
		\node[text width=1cm] at (0.5,-0.7) {0};
		\draw[>=latex, <-, dotted] (0,-0.4) -- (2,-0.4);	
		\draw[>=latex, ->, dotted] (6,-0.4) -- (8,-0.4);
		\node[text width=4cm] at (4.2,-0.4) {Observation window};
	\end{tikzpicture}
	\caption{Example of patients' health trajectories.  Circle represents the observation, and cross represents the death (missing). The end of the observation window is at time $\tau$.\label{extrajectory}}  
\end{figure}

\section{The forward-backward algorithm}
\label{fbal}
Before we discuss inference for the model, we first derive the specifics of the general forward-backward algorithm \citep{baum1966statistical} during the window $\left[0,\tau\right]$ which enable us to simulate the unobserved Markov process $\Xall$. 

In this section, it will be clearer to denote the end of the observation window by $\tau_e$ rather than $\tau$ and to define $X_{\tau_{T+1}}\equiv X_{\tau_e}$. Note that $\tau_{1:T}$ are random observation times but $\tau_e$ is the fixed end of the window. Just as the event $\tau_{1:T}$ is equivalent to having observation-process events at times $\tau_{1:T}$ and at no other times in $[\tau_1,\tau_T]$, the event $\tau_{1:T},\tau_e$ implies, additionally, that there are no observation process events in the interval $(\tau_T,\tau_e]$. Recall, also, that our data format forces us to set $\tau_1=0$, so that the fact that an event happens at time $0$ does not form part of the data; $\tau_1$ now has the same purpose as $\tau_e$, defining the boundary of the observation window. However, since there was an event at time $\tau_1$ there is an observation, $o_1$, which does need to be accounted for.

We use the notation $j:k$ to indicate all indices between and including $j$ and $k$; for example $o_{j:k}\equiv (o_j,o_{j+1},\dots,o_k)$. Thus, we first define the forward variable,
$$
\alpha_{t,k} = \mathbb{P}\left(o_{1:t},\tau_{1:t}, X_{\tau_{t}}=k \right),~t=1,\dots,T,
{
	~~~\mbox{and}~~~
	\alpha_{e,k}
	=
	\mathbb{P}\left(o_{1:T},\tau_{1:T}, \tau_e, X_{\tau_{e}}=k\right)}.
$$

In our case, $\tau_1=0$ is guaranteed, so $\alpha_{1,k}=\nu_k f\left(o_{1} \left| X_{\tau_{1}}=k \right. \right)$. Then for $t=2,\dots,T$,
{
	\begin{equation*}
		\begin{aligned}
			\alpha_{t,k}
			& = \mathbb{P}\left(o_{t}\left| X_{\tau_{t}}=k,o_{1:t-1},\tau_{1:t} \right. \right) \mathbb{P}\left(o_{1:t-1},\tau_{1:t},X_{\tau_{t}}=k \right)\\
			&= f_t\left(o_{t} \left| X_{\tau_{t}}=k \right. \right)  \times \sum \limits_{i=1}^K{\mathbb{P}\left(o_{1:t-1},\tau_{1:t-1}, \tau_t, X_{\tau_{t}}=k, X_{{\tau_{t-1}}}=i \right)}\\
			&= f_t\left(o_t \left| X_{\tau_t}=k \right. \right) 
			\times \sum \limits_{i=1}^K{\mathbb{P}\left(o_{1:t-1}, \tau_{1:t-1}, X_{{\tau_{t-1}}}=i \right)
				\mathbb{P}\left(\tau_t,X_{\tau_{t}}=k\left | X_{{\tau_{t-1}}}=i\right.\right)}\\
			&= f_t\left(o_t \left| X_{\tau_{t}}=k \right. \right) \times
			\sum \limits_{i=1}^K{\alpha_{t-1,i} ~\mathbb{P}\left(\tau_t,X_{\tau_{t}}=k \left | X_{{\tau_{t-1}}}=i\right.\right)}.
		\end{aligned}
\end{equation*}}
{Further, $\alpha_{e,k}=\sum_{i=1}^K\alpha_{T,i}~\mathbb{P}(\tau_e,X_{\tau_e}=k|X_{\tau_T}=i)$.}

To enact the above recursion we must  calculate the quantities, $\mathbb{P}\left(\tau_t,X_{\tau_{t}}=k\left | X_{{\tau_{t-1}}}=i\right.\right)$, $t=2,\dots,T$, and $\mathbb{P}(\tau_e,X_{\tau_e}=k|X_{\tau_T}=i)$. For $t=2,\dots,T$, this is the probability of the next event after time $\tau_{t-1}$ being at time $\tau_t$ and the chain being in state $k$ at this time, all given that the chain was in state $i$ at time $\tau_{t-1}$. The final quantity differs from this in that there has been no event from just after $\tau_T$, up to and including $\tau_e$.

Consider a new process $\{Y_t\}_{t\ge \tau_{t-1}}$ on $\{1,\dots,K,\checkmark\}$. 
The absorbing state $\checkmark$ occurs as soon as there has been at a new event in the Poisson process $N_s$. Specifically, at time $\tau_{t-1}$ the process is in state $i$; \emph{i.e.}, $Y_{\tau_{t-1}}=X_{\tau_{t-1}}$. Up until the next event in the Poisson process, $N_s$, we continue to have $Y_{s}=X_s$, but from the time of the next Poisson event onwards, $Y_{s}=\checkmark$. The process $\{Y_s\}_{s\ge \tau_{t-1}}$ is a CTMC \citep{mark2013algorithm}, and its generator is 
\[
\left(\begin{array}{cc}
	Q-\mathbf{\Lambda} &\blambda^\top\\
	\mathbf{0}^\top & 0\\
\end{array}
\right)
\]
where $\mathbf{\Lambda}=\text{diag}\left(\lambda_1,\ldots,\lambda_K\right)$. Writing $Y_{\tau^-}$ for $\lim_{s\uparrow \tau}Y_s$, and defining $\eta_{t-1,t,i,k}:=\left[\exp\left\{(Q-\mathbf{\Lambda})(\tau_{t}-\tau_{t-1})\right\}\right]_{i,k}$, we see that
\[
\mathbb{P}\left(\tau_t,X_{\tau_{t}}=k\left | X_{{\tau_{t-1}}}=i\right.\right)
=
\mathbb{P}\left(Y_{\tau_t^-}=k\left|Y_{\tau_{t-1}}=i\right.\right)\lambda_k
=
\left[\exp\left\{(Q-\mathbf{\Lambda})(\tau_{t}-\tau_{t-1})\right\}\right]_{i,k}\lambda_k
=
\eta_{t-1,t,i,k}\lambda_k.
\]
{Similarly,
	\[
	\mathbb{P}(X_{\tau_e}=k,\tau_e|X_{\tau_T}=i)
	=
	\left[\exp\left\{(Q-\mathbf{\Lambda})(\tau_{e}-\tau_{T})\right\}\right]_{i,k}
	=:\eta_{T,e,i,k}.
	\]
}
Therefore, the recursions for the forward variable are
$$
\alpha_{t,k}=f_t\left(o_{t} \left| X_{\tau_{t}}=k \right. \right) \times
\sum \limits_{i=1}^K{\alpha_{t-1,i}~\eta_{t-1,t,i,k}\lambda_k}~~~(t=2,\dots,T),$$ 
{with $\alpha_{e,k}=\sum_{i=1}^K\alpha_{T,i}~\eta_{T,e,i,k}$.}

From its definition, $a_{e,k}\propto \mathbb{P}(X_{\tau_e}=k | o_{1:T},\tau_{1:T},\tau_e)$ so we may sample a value for $x_{\tau_e}$ with the probabilities of the individual states proportional to the elements of $a_e$. Suppose that we have simulated $x_{\tau_e},x_{\tau_T},\dots,x_{\tau_{t+1}}$ from their joint distribution given $o_{1:T}$, $\tau_{1:T}$ and $\tau_e$. We now describe how to simulate $x_{\tau_t}$ backward from its conditional distribution given $x_{\tau_e},x_{\tau_T},\dots,x_{\tau_{t+1}}$, $o_{1:T}$, $\tau_{1:T}$ and $\tau_e$. From the hidden-Markov structure, conditional on $x_{\tau_{t+1}}$, all random variables with time indices before $t+1$ are independent of those with indices after $t+1$, so
\[
p_{t,j,k}:=
\mathbb{P}(X_{\tau_t}=j|x_{\tau_e},x_{\tau_T},\dots,x_{\tau_{t+2}},X_{\tau_{t+1}}=k, o_{1:T}, \tau_{1:T},\tau_e)
=
\mathbb{P}(X_{\tau_t}=j|X_{\tau_{t+1}}=k, o_{1:t}, \tau_{1:t+1}),
\]
where $\tau_{T+1}$ is understood to be $\tau_e$. Hence
\[
p_{t,j,k}\propto b_{t,j,k}:=
\mathbb{P}(o_{1:t},\tau_{1:t+1},X_{\tau_t}=j,X_{\tau_{t+1}}=k)
=
\mathbb{P}(o_{1:t},\tau_{1:t},X_{\tau_t}=j)\mathbb{P}(\tau_{t+1},X_{\tau_{t+1}}=k|X_{\tau_t}=j),
\]
again using the conditional independence structure. Thus, for $t=1,\dots,T-1$,
\[
b_{t,j,k} =
\alpha_{t,j} \eta_{t,t+1,j,k}\lambda_k
\propto
\alpha_{t,j} \eta_{t,t+1,j,k},
\]
since $k$ is known. For $t=T$, the constant $\lambda_k$ term does not appear in the first place, leading to the same result. Finally,
\[
\mathbb{P}\left(X_{\tau_{t}}=j \left|x_{\tau_e},x_{\tau_T},\dots,x_{\tau_{t+2}},X_{\tau_{t+1}}=k,o_{1:T},\tau_{1:T},\tau_e \right. \right) =
\frac{b_{t,j,k}}{\sum_{i=1}^K b_{t,i,k}}.
\]

The above recursions enable us to simulate $X_{\tau_t}$, $t=e,T,\dots,1$, conditional on $\bB,\nu,Q,\blambda$,  $o_{1:T}$ and $\tau_{1:T}$ {and (no further observations up to) $\tau_e$}. To simulate from $\Xall$ we also need to fill in the the gaps between the observation times. The conditional dependence structure of the model means that given $X_{\tau_{t-1}}$ and $X_{\tau_t}$, $X_{[\tau_{t-1},\tau_t]}$ is independent of $O_{1:T}$ and of $X_{\tau_1},\dots, X_{\tau_{t-2}}$ and $X_{\tau_{t+1}},\dots,X_{\tau_T}$. Furthermore, given $\tau_{t-1}$ and $\tau_t$, $X_{[\tau_{t-1},\tau_t]}$ is independent of $\tau_{1:T}$. However, the fact that two neighbouring event times, $\tau_{t-1}$ and $\tau_t$ bracket the time interval of interest means that there are no $N_s$ events in $(\tau_{t-1},\tau_t)$.

Combining the above information, to simulate from $X_{(\tau_{t-1},\tau_t)}$ given $o_{1:T},\tau_{1:T}$ and $X_{\tau_1},\dots,X_{\tau_T}$ with $X_{\tau_{t-1}}=j$ and $X_{\tau_t}=k$ we must simulate from $Y_{(\tau_{t-1},\tau_t)}$ given $Y_{\tau_{t-1}}=j$ and $Y_{\tau_{t}}=k$. We employ the uniformization method of \cite{fearnhead2006exact} and \cite{rao2013fast}.

\section{Bayesian inference for the MMPP with an outcome process}
\label{Gibbs}
We are interested in Bayesian analysis of the MMPP with an outcome process. Bayesian inference for CTHMMs has been explored under various likelihood formulation, such as simulation-based \citep{luocthmm2018a} and approximation \citep{williams2019bayesian} methods. To the best of our knowledge, it has never been implemented for an MMPP model with an outcome process under a full Bayesian paradigm. We outline the steps for a Metropolis-within-Gibbs MCMC algorithm to draw posterior samples given suitable prior distributions for the parameters. Our scheme uses the complete data log-likelihood in \eqref{eqlik} in the spirit of \cite{luocthmm2018a} and \cite{fearnhead2006exact}. Throughout, we assume that the number of states, $K$, is fixed and known.

\subsection{Prior distributions}
\label{sec.priors}
As stated in Section \ref{MMPP}, the model for the outcome process when $X_{\tau_t}=k$, $f_k(o_t|z_t,\beta_k)$ is such that a family of conditionally conjugate priors exists; we choose the prior for each $\beta_k$ from this family, and set
$\pi_0(\bB)=\prod_{k=1}^K \pi_{0,k}(\beta_k)$. The initial distribution, $\nu$, is given a Dirichlet prior, so that, conditional on $X_0=x_0$, its posterior is also Dirichlet.

For $\blambda$ and $\bQ$ we choose
\[
\pi_0(\blambda)=\prod_{k=1}^K \mathsf{Gam}(\lambda_k;a^\lambda_k,b^\lambda_k)
~~~\mbox{and}~~~
\pi_0(Q)=\prod_{j=1}^K\prod_{k\ne j}\mathsf{Gam}(q_{j,k};a^q_{j,k},b^q_{j,k}),
\]
where $\mathsf{Gam}(x;a,b)$ is the density of a $\mathsf{Gamma}(a,b)$ random variable, evaluated at $x$. From \eqref{eqlik}, conditional on $\Xall$ and $\tau_{1:T}$, the posterior distributions for each $\lambda_k$ and $q_{j,k}$ ($j\ne k$) are mutually independent and all have Gamma distributions.

For $\blambda$ and $\bQ$, it is tempting to represent the absence of knowledge about each parameter through a prior with a low shape parameter, $a<1$. However, any model with $K$ states includes a model with $K-1$ states, with a particular state, $i$, never visited. In this case, the data, $o_{1:T}$ and $\tau_{1:T}$, provide no further information on $\lambda_i$ and $q_{i,k}$ ($k\ne i$). Thus, the posteriors for all $\blambda$ and $\bQ$  parameters are a mixture including at least one component that is simply the prior. If the prior shape parameter for that parameter is below $1$, the prior density rises to $\infty$ as the parameter tends to $0$, and, because the posterior is a mixture density that includes the prior, the posterior density also rises to $\infty$ as each parameter $\downarrow 0$, which usually does \emph{not} represent our prior belief and, moreover, can lead to poor mixing of the MCMC chain.

\subsection{Exact Gibbs sampler}
\label{sec.GS}
Each step of our Gibbs sampler simulates a parameter or parameter vector, or an aspect of the latent process, $\Xall$, from its conditional distribution given all of the other information, including $\tau_{1:T}$ and $o_{1:T}$. 

The presentation in Sections \ref{MMPP} and \ref{fbal} focused on a single individual. To make the full process clear, let there be $N$ subjects, and let the $n$th have $T^n$ observations $o^n_{1:T^n}$ at times $\tau^n_{1:T^n}$. For subject $n$, we shift time so that {$\tau^n_1=0$, set $\tau^n=\tau-\tau^n_{1}$}, where $\tau$ corresponds to the end of 2014, and define the latent path to be $\Xalln$. Each latent path $\Xalln$ gives rise to summary statistics $N_{l,m}^n$, $R_l^n$ and $N_i^n$ as defined for a specific subject in Section \ref{lik}. 
We write $\bTheta=(\bB,\blambda,\bQ,\nu)$ and $\mathcal{D}^n=\{T^n,\tau^n_{1:T^n},O^n_{1:T^n}\}$.

Following \cite{fearnhead2006exact,luocthmm2018a}, the Gibbs sampler has the following distinct steps.

\begin{enumerate}
	\item For each $n$ in $1,\dots,N$, simulate $(x^n_{\tau^n_1},\dots,x^n_{\tau^n_{T^n}}{,x^n_{\tau^n}})$ from its conditional distribution given $\bTheta$, $\mathcal{D}^n$ via the steps in Section \ref{fbal}.
	\item For each $n$ in $1,\dots,N$ and each $t$ in $1,\dots,T^n$, simulate $x_{(\tau^n_{t},\tau^n_{t+1})}$ {(identifying $\tau^n_{T+1}$ with $\tau^n$)} via the direct-sampling method of \cite{hobolth2009simulation} or uniformization in \cite{fearnhead2006exact} and \cite{rao2013fast}. This provides $\xalln$ and, hence, the summaries $R_l^n$, $N_{l,m}^n$ and $N_l^n$, $n=1,\dots,N$.
	\item Update $\nu$ from its conditional Dirichlet posterior given $x_0^{1:N}$.
	\item For $l=1,\dots,K$ and $m\ne l$, update $q_{l,m}$ from its conditional Gamma posterior given $R_l^{1:N}$ and $N_{l,m}^{1:N}$. 
	\item For $k=1,\dots,K$, update each  $\lambda_k$ from its conditional Gamma posterior given $R_k^{1:N}$ and $N_k^{1:N}$.
	\item For $k=1,\dots,K$, update $\beta_k$ from its conditional posterior given $x^1_{\tau^1_1},\dots,x^1_{\tau^1_{T^1}},\dots,x^N_{\tau^N_1},\dots,x^N_{\tau^N_{T^N}}$. 
\end{enumerate}

Our first simulation study involves a two-state chain and allows us to demonstrate the utility of both forms of information: the event times and values observed at these times. Our second simulated example, and our real-data example consider a three-state chain, with the third state being death and where the state can only increase. The  non-zero transition rates are $q_{1,2}$, $q_{1,3}$ and $q_{2,3}$, and the non-zero observation-process rates are $\lambda_1$ and $\lambda_2$.

\section{Empirical studies}
\label{emprical}
\subsection{Example 1}
In this example, we demonstrate the performance of the proposed Gibbs sampler, and investigate the convergence, mixing and inferences on parameters under different scenarios. 

For each of $N=50$ subjects, we simulate a realization of a two-state latent process without the death, $X_{[0,5]}$, using a generator of
\begin{equation*}
	\bQ=\left(\begin{array}{rr}
		-1 & 1\\
		3&-3 \\
	\end{array}
	\right),
\end{equation*}
and an initial distribution of $\nu= (0.8,0.2)$.
The outcome process is  Gaussian, $O_t\sim \mathsf{N}(\beta_k,1)$, where $\beta_k$ is a scalar, so that $\bB=(\beta_1,\beta_2)$. For each individual, the time window is $[0,5]$; individuals are started from the initial distribution, $\nu$, and the trajectory, events and observations simulated for the entire window. Thus, here, in contrast with our real-data scenario, there is no observation at time $0$, and the first forward variable is $\alpha_{0,k}=\nu_k$.

We consider four scenarios in this example.  
\begin{itemize}
	\item Scenario I: $\bB=(-1,1)$ and the point process rate $\lambda=(4,12)$. 
	\item Scenario II:$\bB=(-1,1)$ and the point process rate $\lambda=(8,8)$. 
	\item Scenario III: $\bB=(0.8,1)$ and the point process rate $\lambda=(4,12)$.
	\item Scenario IV: We use the same $\bB$ and $\lambda$ as Scenario III but perform Bayesian inference using a CTHMM \citep{luocthmm2018a}, ignoring the point process.
\end{itemize}
In Scenario I, both the Poisson intensity and the Gaussian expectation are distinct between states, whereas in Scenario II, while the Gaussian expectations are distinct, the Poisson intensities are the same. In Scenario III, the Gaussian expectations are similar but the intensity rates distinct,  while Scenario IV is extremely challenging, with no information from the point process and very little information from the outcome process. 

We place non-informative priors on all the parameters in this study. Each of $\lambda_1$, $\lambda_2$, $q_{12}$ and $q_{21}$ have independent $\mathsf{Gamma}(1,1/8)$ priors.  We place a $\mathcal{N}(0,10000)$ prior on each element of $\bB$, and a $\mathsf{Beta}(1,1)$ prior for $\nu$.

\begin{figure}[ht]
	\centering
	\includegraphics[scale=0.6]{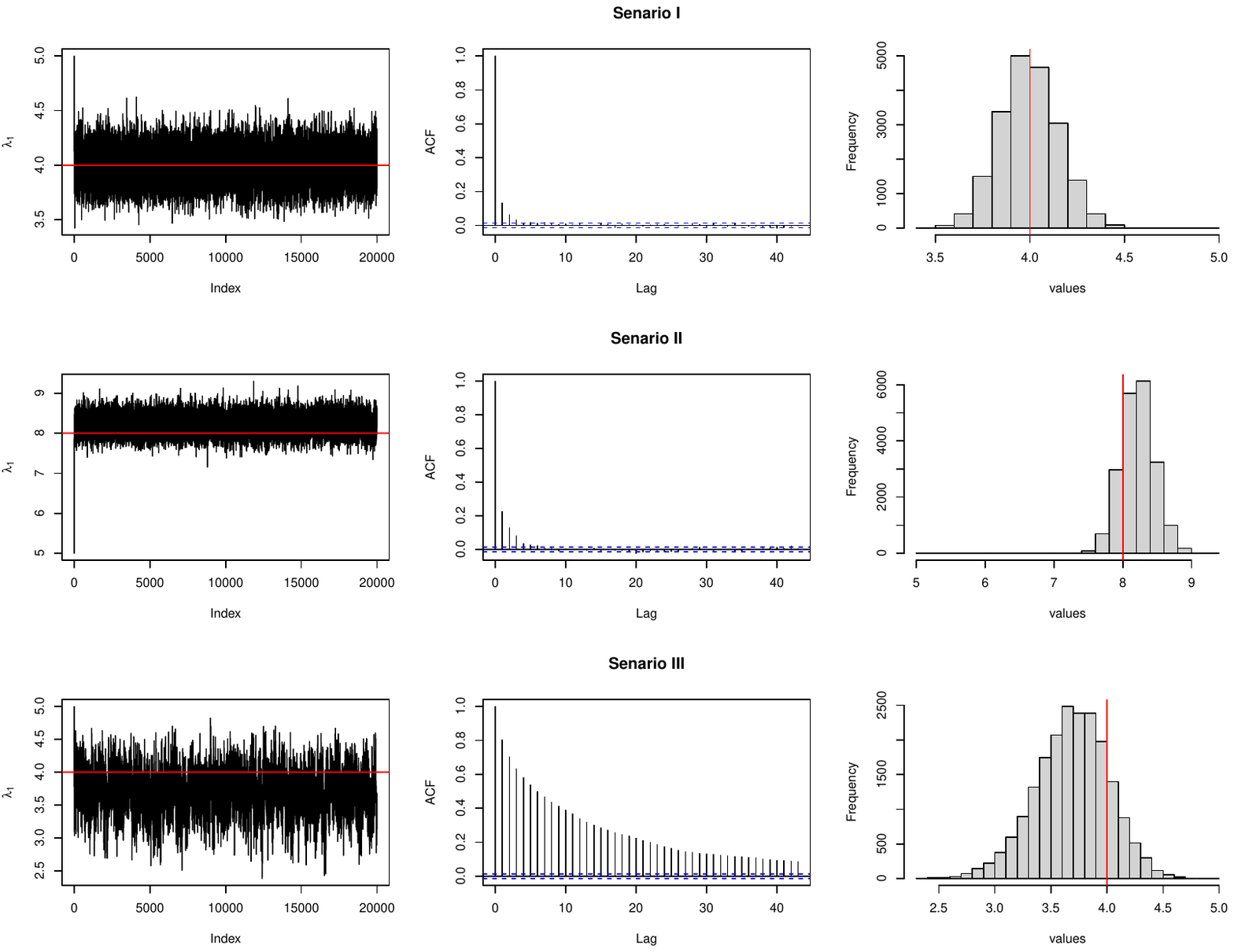}
	\caption{\label{toy1} Example 1: Trace, ACF plots and histogram for $\lambda$ using the exact Gibbs sampler for different scenarios.}
\end{figure}

\begin{figure}[ht]
	\centering
	\includegraphics[scale=0.6]{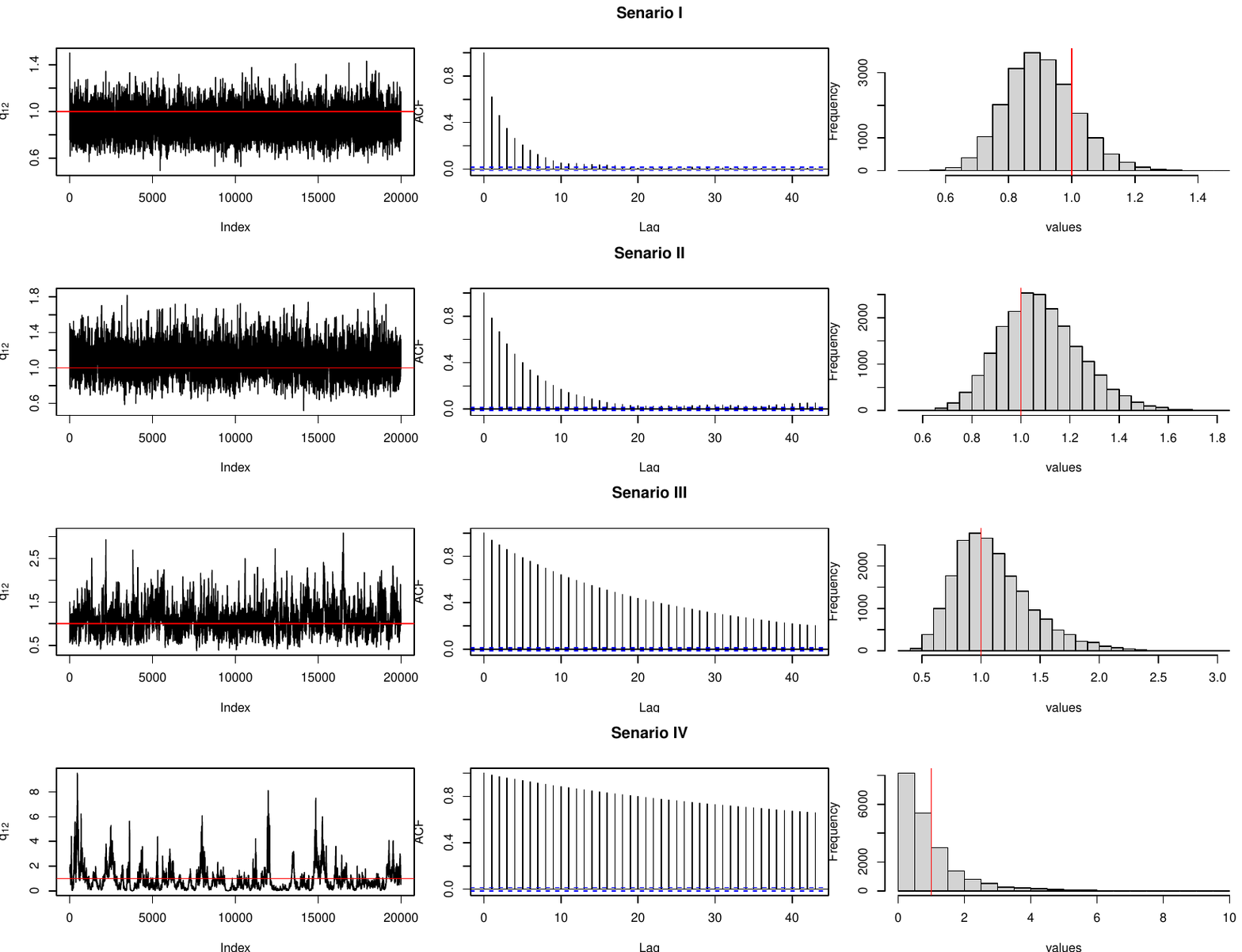}
	\caption{\label{toy2} Example 1:  Trace, ACF plots and histogram for $q_{12}$  using the exact Gibbs sampler for different scenarios.}
\end{figure}

Trace plots, auto-correlations plots and histograms for $\lambda_1$ and $q_{12}$ are shown in Figures \ref{toy1} and \ref{toy2}; Integrated auto correlation times (IACTs) for $\lambda_1$, $\lambda_2$, $q_{1,2}$, $q_{2,1}$, $\beta_1$ and $\beta_2$ are provided in Table \ref{table.IACTa}.  In Scenario I, where the data are generated from a model characterized by clear separation between the states, the Gibbs sampler performs well, with fast convergence and mixing.  In Scenario II, where the timing of the events provides no information about the state, the Gibbs sampler's performance declines slightly, though it still remains promising. The posterior distribution of $\lambda_1$ is slightly off the true value.  Challenges arise when the data are generated from a less well-separated outcome process, leading to notable auto-correlations in the posterior samples, particularly those associated with $\bQ$. The reduction in mixing efficiency is considerable; however, the chain does still mix adequately over the 20,000 iterations shown. By contrast, in Scenario IV, when the point-process information is removed from Scenario III, the exploration of the posterior over the 20,000 iterations is so poor that we would judge that the chain has perhaps not yet even converged. Moreover, the posterior for $q_{1,2}$ includes values that are not strongly supported by the data. This illustrates the most important point: especially in cases where the outcome process is not particularly informative, the information from the timings of the events provides an important contribution to the posterior and, possibly because of the tighter posterior and more clearly defined paths for each $\Xall$, improves the mixing of the chain.

\begin{table}
	\begin{center}
		\begin{tabular}{l|rr|rr|rr}
			Scenario&$\lambda_1$&$\lambda_2$&$q_{1,2}$&$q_{2,1}$&$\beta_{1,1}$&$\beta_{1,2}$\\
			\hline
			I&3.4&5.0&7.8&8.4& 5.4& 4.6\\
			II&3.9&4.8&12.4&10.7&7.1&7.1\\
			III&27.4&22.9&52.7&47.7&13.3&14.5\\
			IV&n/a&n/a&(257.0)&(261.4) & (276.2)& (384.0)\\
			\hline
		\end{tabular}
		\caption{\label{table.IACTa}Example 1: Integrated auto-correlation times (IACTs) of the posterior sample for each parameter in each scenario. Bracketed terms indicate that the effective sample size was less than $100$, so there is reduced confidence in the estimate of the IACT.}
	\end{center}
\end{table}

\subsection{Example 2}
\label{ex1}
In this example, we demonstrate valid inference for simulated data with a similar pattern to our real data. We generate the data from a three-state model which is similar to that used for our real data, with  $N=500$ subjects. For each subject, we simulate a realization of a three-state latent process with the death as the absorbing state, $X_{[0,10]}$,
\begin{equation*}
	\bQ=\left(\begin{array}{rrrr}
		-0.21 & 0.20&  0.01\\
		0.00 & -0.05  &0.05\\
		0.00&  0.00 & 0.00  \\
	\end{array}
	\right),
\end{equation*}
so trajectories can only progress forward with no chances to transit back to previous states. The outcome process is $O_t\sim \mathsf{Poisson}(\mu_t)$ with $\log \mu_t = B_{1,k}+z_t B_{2,k}$ when in state $k$, and where the time-varying covariate is $Z_t\sim$Bernoulli$\left(0.65\right)$; we set
\begin{equation*}
	\bB=\left(\begin{array}{rr}
		-0.69 & 0.77 \\
		-0.13 & -0.39  \\
	\end{array}
	\right).
\end{equation*}
This is equivalent to $O_t|(Z_t=1)\sim \mathsf{Poisson}(\mu_{1,k})$ and  $O_t|(Z_t=0)\sim \mathsf{Poisson}(\mu_{0,k})$, with $\log \mu_{1,k}=B_{1,k}+B_{2,k}$ and $\log \mu_{0,k}=B_{1,k}$. In this case, the outcome expectations (marginalized over the covariate) are $0.35 \mu_{0,k}+0.65\mu_{1,k}$, which are $0.46$ and $1.71$ for the two states, respectively.
The point process rate is set as $\lambda= (6,10,0)^\top$. State 3 is the death state without any observed outcome. If the patient progresses to the absorbing state before the observation window, i.e. death, then the last observation is the final one before death. Otherwise, the last  observation is the one right before the observation window, $\tau=10$.  We assumed that the first observation is made at time 0 according to the initial distribution. Similar to Example 1,  we place non-informative priors on all the parameters in this study. Each of $\lambda_1$, $\lambda_2$, $q_{12}$ and $q_{21}$ have independent $\mathsf{Gamma}(1,1/8)$ priors.  We place independent $\mathsf{Gamma}(0.1,0.1)$ priors on each of the four Poisson expectations, $\mu_{i,k}$, $i=0,1$, $k=1,2$, and a $\mathsf{Beta}(1,1)$ prior for $\nu_1$. 

\begin{figure}[ht]
	\centering
	\includegraphics[scale=0.4]{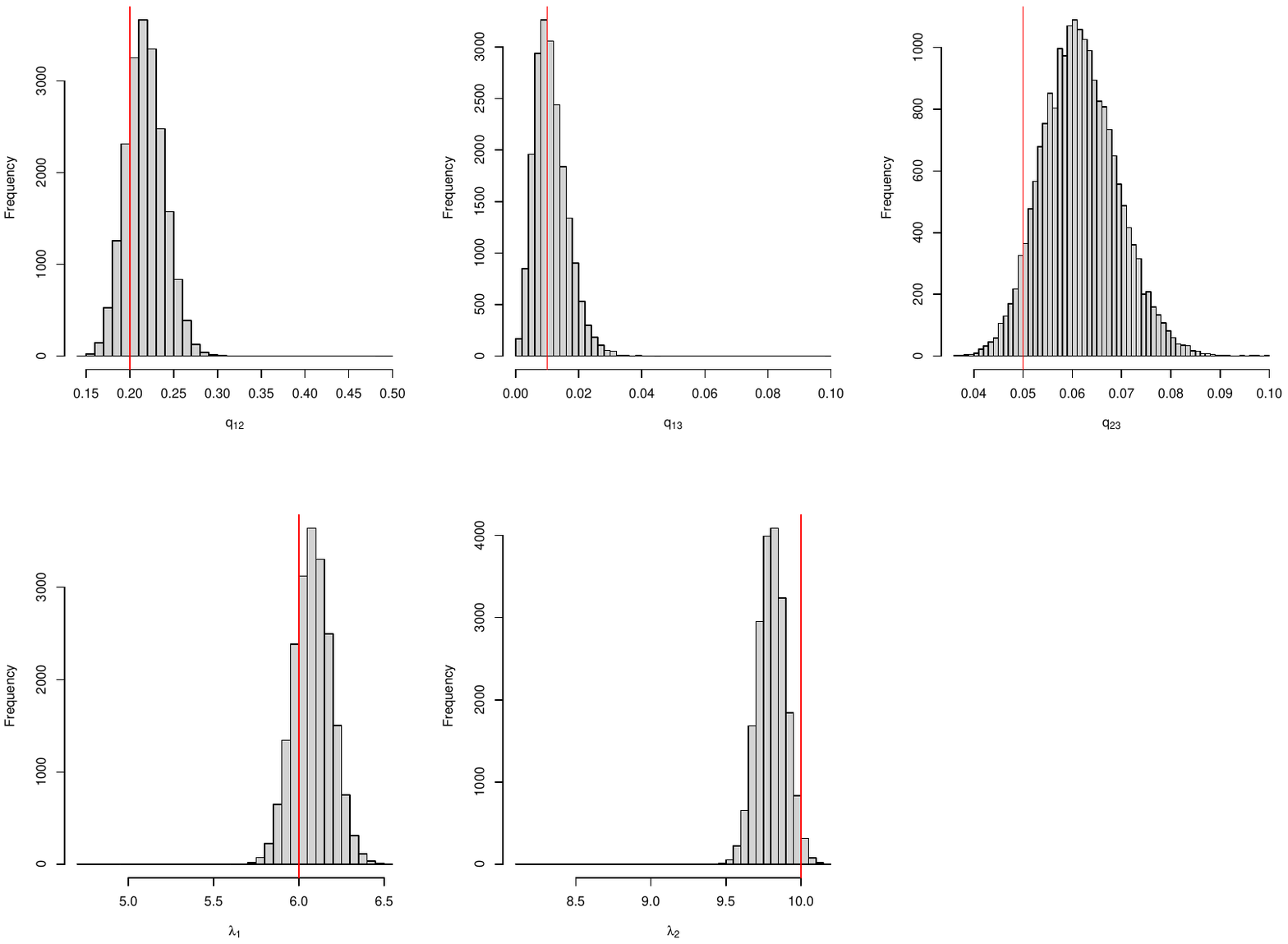}
	\caption{\label{ex2:2}Example 2: marginal histograms of $q_{12}$, $q_{1,3}$, $q_{2,3}$, $\lambda_1$ and $\lambda_2$; the associated vertical red lines depict the true parameter value.}
\end{figure}

Table \ref{table.IACTaex2} presents the integrated auto correlation times (IACTs) for $\blambda$, $\bQ$ and $\bB$ over 20,000 iterations and shows that all of these parameters mix well. Figure \ref{ex2:2} presents a histogram of the marginal posterior for each of these parameters, with each true parameter value falling within the main posterior mass. This example demonstrate that by simulating the data-induced window similar to the real data, we can achieve valid inference.
\begin{table}
	\begin{center}
		\begin{tabular}{rr|rrr|rrrr}
			\hline
			$\lambda_1$&$\lambda_2$&$q_{1,2}$&$q_{1,3}$ & $q_{2,3}$& $\beta_{1,1}$& $\beta_{1,2}$ & $\beta_{2,1}$ & $\beta_{2,2}$\\
			\hline
			3.0&3.0&3.2&3.5&3.1&3.4&3.7&3.2&3.0\\
			\hline
		\end{tabular}
		\caption{\label{table.IACTaex2} Example 2: Integrated auto correlation times (IACTs) of the posterior sample for each parameter.}
	\end{center}
\end{table}

\subsection{Application: Canadian healthcare system data}
In this section, we delve into the proposed methodology designed to enhance our understanding of the progression of chronic diseases, particularly COPD. The significance of this understanding lies in its potential to help  with effective management of healthcare systems and pave the way for methods of early diagnosis and personalized care, as the model can help the management team understand the ``care path" or ``trajectory" of the cohort before they become severely ill. While clinical and administrative data harbor valuable insights, traditional methods for modelling disease progression face limitations when dealing with data collected at irregular intervals and interaction frequency, such as when patients choose to engage with the healthcare system.

To address these challenges, our proposed methodology is applied to a COPD cohort derived from healthcare administrative data. An open, dynamic cohort was established, comprising a 25\% random sample of Montreal residents in Canada. In 1998, a similar random sample was drawn from the R\'{e}gie de l'assurance maladie du Qu\'{e}bec (RAMQ)-registered population within the census metropolitan area of Montreal. Subsequently, at the onset of each subsequent year, 25\% of individuals born in Montreal or relocating to Montreal within the past year were sampled to maintain cohort representativeness. Follow-up concluded upon death or a change in residential address outside Montreal; however, neither death nor the time of address change is recorded in our data. This administrative database encompasses outpatient diagnoses and procedures submitted through RAMQ billing claims, as well as inpatient claims for procedures and diagnoses. Drug dispensing data are available for individuals covered by drug insurance through RAMQ, which includes roughly half the population, especially all those aged 65 and above. All data are interconnected through an anonymized version of the RAMQ number, and administered by the Surveillance Lab at McGill Clinical \& Health Informatics research group.

\begin{figure}[ht]
	\centering
	\includegraphics[scale=0.5]{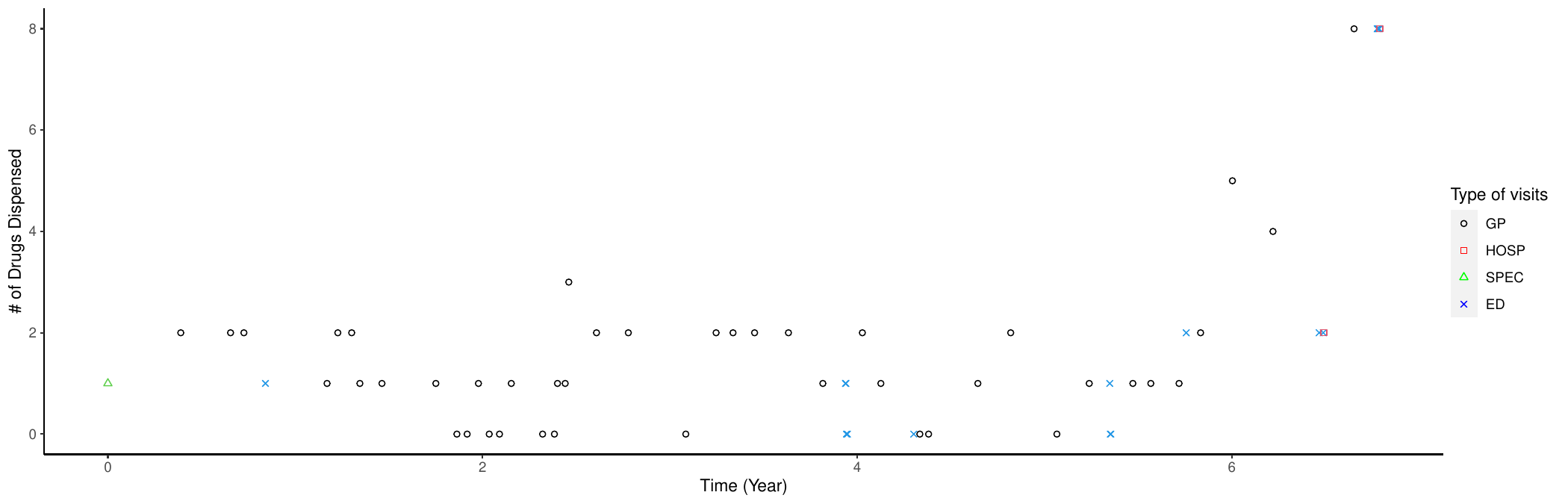}
	\caption{\label{ex3:1}COPD patient's trajectory from the Canadian healthcare system data.}
\end{figure}

Utilizing established case definitions based on diagnostic codes \citep{lix2018canadian}, a total of 76,888 COPD patients were enrolled, with an incident event (ICD-9 491x, 492x, 496x; ICD-10 J41-J44) occurring after a minimum of two years at risk with no events. The observation period spanned from January 1998, commencing with the patients' first diagnosis, until December 2014. Physicians observed these patients solely during medical visits, which transpired when patients chose to interact with the healthcare system. In these medical visits, pertinent medical information was gathered, and from this observed information, we aim to infer the patients' unobserved disease status, which is represented using a discrete disease state model. In this study, the disease status was indirectly gauged through a proxy: the number of prescribed medications at the time of observation. Notably, this information was available exclusively for patients with drug insurance, leading us to focus on patients aged over 65 years, constituting the study population for our analysis. This full dataset has been previously analyzed in \cite{luocthmm2018a}. Due to the data restriction, we are allowed to randomly sample 1,000 patients to demonstrate our proposed method. Figure \ref{ex3:1} shows one COPD patient's follow up from this dataset. As shown in the figure, we observe the number of drugs prescribed as the observation process. In addition, we have also observe variations in the type of encounters, \emph{i.e.}, general practitioner visit (GP), hospitalization (HOSP), specialist visit (SPEC) and emergency department visit (ED), which will be incorporated as time-varying covariates in our model. These changes in the observation point process, transitioning from sparse to dense, reflect the evolving severity of the chronic disease under study. In the dataset, the average number of drugs prescribed from a visit for GP, ED, HOSP, and SPEC is 5.06, 5.57, 5.28, and 5.17 respectively, suggesting little deviation among the types of healthcare utilization. Over the 1,000 patients, the average time from the first recorded interaction until the last observation is about 13.71 years.

We apply the model from Section \ref{MMPP} to this sample dataset, with the inference methodology from Section \ref{Gibbs}. We focus on a three-state model with State 3 (the unobserved death state) as an absorbing state. We place independent $\mathsf{Gamma}(1,1/8)$ priors on each $\lambda_i$ and all elements in $\bQ$, and a $\mathsf{Gamma}(0.1,0.1)$ prior on each expected number of drugs dispensed for each healthcare utilization,  and a $\mathsf{Beta}(1,1)$ prior for $(\nu_1,\nu_2)$, with $\nu_3=0$. The outcome  process is modeled with a Poisson distribution, with the types of health care utilization (HOSP, SPEC, GP, and ED) as a time-dependent covariate; equivalently, for each of states 1 and 2 there are four parameters, indicating the expected number of drugs dispensed for each of the four values of the covariate. 

We ran the algorithm for 20,000 iterations; for all parameters, the estimated IACTs were less than 5, indicating excellent mixing of the MCMC algorithm. Table \ref{real} displays the estimates for all of the model parameters. The narrow 95\% credible intervals associated with these estimated parameters indicate low uncertainty within the estimated densities. The posterior median for the number of prescribed drugs rises by about a factor of four from State 1 to State 2, reflecting the worsening severity of patients' health conditions. In State 2, expected drug dispensations for emergency department visits and hospitalizations are slightly higher than other types of visits. Point process rates are roughly 35\% higher in State 2 than in State 1, suggesting increased healthcare system interactions during poorer health conditions. This indicates a correlation between disease severity and healthcare utilization, with state 2 indicating the higher level of healthcare engagement among patients. In the state transition dynamics, the median stay in State 1 is roughly $2.3$ years and the probability that a patient is still in State 1 after four years is approximately $0.30$. When a transition from state 1 occurs, the probability that it will be directly to the death state is $0.23$. 
The estimates for state 2 suggest a more persistent state (median stay being over 4 years) and potentially more severe state of illness. Patients in State 1 are less likely to die (or move from the province) in any given year, compared to those in State 2. The median time to death in State 2 is 4.3 years.  We also notice that $\nu_1\approx\nu_2$ meaning that conditional on a healthcare visit, patients are almost equally likely to start in each of the two states in the heathcare system. 

We also ran the Gibbs sampler without the point process as with Scenario IV in Example 1. The parameter estimates are reproduced in the Supplementary Material. They show that ignoring the point process information makes patients seem to  remain in State 1 for  longer. The other substantial difference is that the transition rate to the death state from State 1 is slightly greater than that from State 2. We also fitted a four-state model, again with only transitions to higher states allowed, and now with State 4 (the unobserved death state) as an absorbing state. We compared a Laplace-based  approximation of its marginal likelihood with a similar approximation to the marginal likelihood of the three-state model; see the Supplementary Material for details.  The estimated marginal likelihoods for three-state and four-state models are $-2702.93$ and $-2687.05$ respectively, leading to corresponding BICs of $5502.57, 5612.19$. This shows that the three-state model is preferable; results for the four-state model are also presented in the Supplementary Material.

This inferred state dynamic provides valuable insights for healthcare management, offering guidance on the performance of current COPD interventions. The tendency for patients in State 1 to remain in that state initially before transitioning to State 2 underscores the need for interventions aimed at preventing or delaying disease progression, particularly during the early stages. For patients already in State 2, the observed persistence suggests a potentially more challenging treatment landscape, emphasizing the importance of comprehensive and long-term management strategies. Understanding these state transition dynamics informs healthcare providers and policymakers about the effectiveness of existing interventions and guides the development of targeted strategies to improve patient outcomes. By identifying critical transition points and patient populations at higher risk of disease progression, healthcare resources can be allocated more effectively to provide timely and appropriate interventions, ultimately improving the management and care of individuals with COPD. However, it is important to note that there is no guarantee that the latent state and trajectory groupings will directly correspond to the underlying biologic processes or disease stages as our initial model is built without clinical measurements. Therefore, the identified states correspond to the general health conditions of COPD patients, implying that a higher disease level or intensity of healthcare encounters does not strictly indicate a deterioration in COPD but could signify any temporary or prolonged period of poor health condition. By using this Canadian cohort dataset, the initial latent groupings are expected to generate hypotheses about how the patients are being managed. 

\begin{table}
	\caption{\label{real} Posterior median of MMPP parameters associated with 95\% credible intervals.}
	\centering
	{\begin{tabular*}{40pc}{@{\hskip5pt}@{\extracolsep{\fill}}c@{}c@{}c@{\hskip5pt}}
			\hline
			Parameter  & State 1 &  State 2\\
			\hline
			$\lambda$ (Marked point process rate) &   5.07 (4.97, 5.16)& 6.82 (6.72, 6.93) \\
			$\nu$ (initial state distribution)  & 0.54 (0.51, 0.58) & 0.46 (0.42, 0.49)  \\
			Expected number of drugs dispensed (GP) & 2.12 (2.10, 2.14) & 7.28 (7.26, 7.31) \\
			Expected number of drugs dispensed (ED)& 1.72 (1.67, 1.78) & 7.47 (7.43, 7.53)  \\
			Expected number of drugs dispensed (HOSP)&  1.77 (1.72, 1.81) &7.36 (7.31, 7.41) \\
			Expected number of drugs dispensed (SPEC)&  1.79 (1.75, 1.85) & 7.16 (7.10, 7.21)   \\
			$q_{1.}$ (Transition rate from State 1) &$-$&0.23 (0.21, 0.25) \\
			$q_{.3}$ (Transition rate to death state) &0.07 (0.06, 0.08) &0.16 (0.14, 0.17) \\
			
			\hline
		\end{tabular*}
	}
\end{table}

\section{Discussion}
\label{dis}
In this paper, we have considered a generalization of both the hidden Markov model and the MMPP to an MMPP with an outcome process,  and we have proposed an exact Gibbs sampler to facilitate Bayesian inference. Our data model also incorporates an unobserved terminating event (death or moving away from the area). 

This advance is driven by the need to analyze longitudinal data and model an unobserved death so as to comprehend the progression of a disease processes. A critical decision revolves around whether observation times are considered non-informative or informative, on the state of the Markov chain, which relates to the disease severity. Neglecting to account for an informative observation process in the disease dynamics or to incorporate an unobserved terminating event, such as death, can potentially lead to biased parameter estimates.  With the large amount of claim data and healthcare data available, there is a pressing need to work on appropriate statistical modelling framework for longitudinal healthcare studies. The proposed technique for inferring death prevents the bias that would arise if a model without death were employed with a shortened observation window only up to the final event time. Our approach offers a framework for analyzing complex healthcare data, providing insights into disease progression and patient health/death states that can inform clinical decision-making and healthcare interventions. 

Since our algorithm is a Gibbs sampler, it requires no tuning parameters and could be used ``off the shelf" by practitioners.  Simulation studies indicate the effectiveness of the Gibbs sampler and show how the event times contribute vital information, especially when the outcome process lacks a clear separation between the states.  By integrating observation times and death into the model, we enhance our understanding of disease dynamics and healthcare utilization patterns, thus advancing the methodology for analyzing longitudinal data in healthcare research. In our application, we imposed a constraint that the process can only progress forward since COPD gets progressively worse over time \citep{COPD}. However, one could make no restrictions on the allowed transitions in the state space. In the absence of other information, the hidden state would then be non-identifiable with all permutations of the state labels equivalent. To avoid the ensuing  identifiability issue, meaningful priors could be placed on the outcome models for each state or the outcome process event rates – that is, we would distinguish the states via their impact on the observable quantities or imposing certain restrictions on $\bQ$ (see Section 4 in \citealp{luocthmm2018a} for a detailed discussion).

Although the proposed model possesses a flexible structure, there remains some scope for future research to explore several possibilities in analyzing such irregularly spaced longitudinal data.   
The current setup assumes time-homogeneous processes $X_s$ and $N_s$. Both processes may plausibly depend on time. For instance, within any given health state, older patients might engage more frequently with the healthcare system. However, introducing complex structures to model the transition in $X_s$ poses computational challenges in both frequentist and Bayesian inference \citep[see for example,][and a discussion in the Supplementary Material of this article]{kendall2024beyond}. In addition, the issue of identifying the number of hidden states in MMPPs has been a long-standing challenge. In the context of HMMs, this topic has been explored using approaches such as marginal likelihood, cross-validated likelihood, and complete-data likelihood \citep[see,][]{celeux2008selecting,pohle2017selecting,luo2021bayesian}; however, model selection criteria based on these quantities become unreliable as the dimensionality increases beyond 50 \citep{chen2008extended}. Moreover, while the current model categorizes patients into discrete states, which is appropriate for some chronic diseases thought to have discrete stages of progression, relaxing the assumption of discrete states could yield insights. Relaxing the assumption that patients evolve through discrete states of underlying disease progression, would allow estimation of a continuum of severity that may better represent the trajectory of COPD and other chronic diseases. Substantial advances in statistical methodology would be required to extend this approach to a continuous state space and to allow the latent state distribution to have an interpretation as a continuous index or score. For example, one could use the features included in comorbidity indices to measure multimorbidity in terms of the ability to predict future mortality and health services use. The inference problem for this type of assumption involves multivariate diffusion processes, using discrete-time data that may be incomplete and affected by measurement errors \citep[e.g.,][]{golightly2022augmented}.

\section*{Acknowledgments}
The authors would like to thank Prof. David Buckeridge from the Surveillance Lab at McGill Clinical \& Health Informatics research group for granting us access to the real data that enabled us to demonstrate the empirical study in this paper.

\clearpage

\appendix

\begin{center}
	{\LARGE\bf Supplementary Materials for ``Bayesian inference for the Markov-modulated Poisson process with an outcome process"}
\end{center}

\section{Using the Metropolis--Hastings algorithm for inference}

We have presented an exact Gibbs sampler for inference on a Markov modulated Poisson process with an observation process at event times. If non-conjugate priors were used, one would need to use the Metropolis--Hastings algorithm. The Gibbs sampler has two appealing properties compared with the Metropolis--Hastings algorithm.

Firstly, it is automatic: the user can simply apply it to their data and the algorithm will run straight ``out of the box". By contrast, the Metropolis--Hastings algorithm requires, firstly a choice of blocking - should all the $\boldsymbol{\beta}$ parameters be updated simultaneously then all of the $\mathbf{Q}$ parameters and then the vector $\boldsymbol{\lambda}$, should all be updated at once or is some other choice preferable? Then, for the commonly used random walk proposal, for example, proposal scalings and variance matrices need to be chosen. These either require initial tuning runs or some form of adaptive MCMC \cite[e.g.][]{RobRos2009} which in turn needs a choice of hyperparameters and careful watching. 

Secondly, sampling from the conditional posterior explores the \emph{whole} of the conditional state space for the parameter and with guaranteed acceptance. For example, with Gaussian observations, the covariate vector $\boldsymbol{\beta}$ is sampled directly from its conditional posterior whatever the dimension, $\mathsf{dim}(\boldsymbol{\beta})$. By contrast with a random walk Metropolis proposal the magnitude of the proposed jump in $\boldsymbol{\beta}$ must shrink in proportion to $1/\sqrt{\mathsf{dim}(\boldsymbol{\beta})}$ \cite[e.g.][]{RobRos2001}. The Metropolis-adjusted Langevin algorithm and Hamiltonian Monte Carlo have better scaling, but both require the computation of gradients of the log posterior.

\section{Non-homogeneous transitions}

Inherent in the use of a time-homogeneous hidden Markov model is that the waiting times between the changes of state of the hidden process are exponentially distributed and that the probability of a particular change given that a state change has occurred does not depend on the amount of time already spent in the current state. However, it is possible that this is not the case in reality. We explain why inference using both the timing of the events, $\tau_1,\tau_2,\dots$ and the observations $o_1,o_2,\dots$ is difficult in the general scenario, but then illustrate a particular case where a natural extension of our algorithm is possible: when waiting times have an Erlang distribution.

It is helpful to re-examine why relatively tractable inference, as described in Sections 3 and 4, has been possible. Transitions of the underlying Markov model occur according to constant-rate Poisson processes. For example, in a chain with two live states and a death state. If $X_0=1$ then there are two Poisson processes, one with rate $q_{1,2}$ and one with rate $q_{1,3}$, and whichever has the first  event determines the next change of state. Crucially, we have also modeled the events in the observation process as arising from a Poisson process; in our running example, since $X_0=1$, this has a rate of $\lambda_1$. The fact that both transitions \emph{and} observation events occurred according to Poisson processes allowed us to consider an extended state space and write down the generator in Section 3:
$$
\begin{pmatrix}
	\mathbf{Q}-\boldsymbol{\Lambda}&\boldsymbol{\lambda}\\
	\mathbf{0}&0
\end{pmatrix}.
$$
Exponentiating the generator produced transition probabilities that enabled us to employ the forward-backward algorithm to simulate the states of the hidden chain at the event times of the observation process. Maximum-likelihood and Bayesian inference for homogeneous hidden Markov models is  covered in depth in \cite{CapMouRyd2005} and there have been many advances for the homogeneous case since. Methods for Bayesian inference on the Markov modulated Poisson process are described in \cite{fearnhead2006exact,rao2013fast}; maximum-likelihood inference on a combined homogeneous HMM with an informative observation process is performed via the EM algorithm in \cite{lange2015joint} and \cite{mews2023markov}. To the best of our knowledge this article is the first to address Bayesian inference the combination.  

One natural mechanism for allowing for different waiting distributions and time-dependent transition probabilities is by making the Poisson processes for the transitions non-homogeneous. 
There has been work in the sphere of inference for non-parametric Poisson intensities for a general non-Markovian model where the states themselves are observed \cite[]{Titman2011, kendall2024beyond, PutterSPitoni2018}, maximum-likelihood for non-homogeneous hidden Markov models using numerical integration \cite[]{TitmanRnhm},  and Bayesian estimation for a Poisson process with a non-parametric intensity  \cite[]{AdamsMurrayMacKay2009,Alie2023}. Whilst the general technique of particle MCMC \cite[]{AndDouHol2010} can always be employed as long as it is possible to simulate from the hidden process, it involves further choices such as the number of particles and the type of particle filter to be used. We would like to extend our relatively straightforward methodology to the non-homogeneous case, but there is no fixed rate matrix to exponentiate and so the  transition probabilities between event times of the observation process are intractable \citep[e.g.,][]{kendall2024beyond}; moreover there would, in general, be no conjugate prior, so parameter inference would require the Metropolis--Hastings algorithm.  We, therefore, see no natural extension of our method to the general case. However, it is possible to extend our method to the special case of a model where the timings for the potential state changes have Erlang distributions. 

An Erlang distribution is a Gamma distribution with an integer shape parameter, $k$, and a rate parameter $q$. Hence, an $\mathsf{Erlang}(k,q)$ random variable is a sum of $k$ independent $\mathsf{Exp}(q)$ random variables. Transitions can, therefore be modeling by adding $k-1$ interim states for each actual state transition. We illustrate this by considering the three-state model from Section 5, where State 3 corresponds to death and where there are no transitions back to lower-numbered states. We allow for a single interim transition for each state change and label the states with the partial transitions that have occurred in brackets. We also allow the first partial transition towards death to occur from either State 1 or State 2, so that the marginal time to death does not have an Erlang distribution unless $q_{1,3}=q_{2,3}$; we note that in the Markov chain in Section 5, the marginal time to death does not have an exponential distribution unless $q_{1,3}=q_{2,3}$. The generator is
$$
\mathbf{Q}=
\begin{array}{r|ccccccc|}
	&1&1(2)&1(3)&1(2,3)&2&2(3)&3\\
	\hline
	1&-&q_{1,2}&q_{1,3}&0&0&0&0\\
	1(2)&0&-&0&q_{1,3}&q_{1,2}&0&0\\
	1(3)&0&0&-&q_{1,2}&0&0&q_{1,3}\\
	1(2,3)&0&0&0&-&0&q_{1,2}&q_{1,3}\\
	2&0&0&0&0&-&q_{1,3}&0\\
	2(3)&0&0&0&0&0&-&q_{2,3}\\
	3&0&0&0&0&0&0&0
\end{array}~
,
$$
where for simplicity of presentation, the minus sign, $-$, in each row stands for the negative quantity that will make the entries in the row sum to $0$. The corresponding vector for the rate of the observation process is $\boldsymbol{\lambda}=[\lambda_1,\lambda_1,\lambda_1,\lambda_1,\lambda_2,\lambda_2,0]^\top$.

\section{Continuous-path versus Discrete-time Likelihoods}
In a CTHMM scenario it is possible to integrate out the hidden Markov chain between observation times to give the likelihood
\[
\mathcal{L}(\mathbf{Q};x_{[0,\tau]})=
\prod_{t=1}^T
\exp\left\{\mathbf{Q}\Delta_t\right\}_{x_{\tau_t},x_{\tau_{t+1}}},
\]
where $\Delta_t=\tau_{t+1}-\tau_t$ and $\tau_{T+1}=\tau_e$. That is, we impute the states at each observation time but we do not fill in the full continuous-time Markov chain between states. In our setting, this could be adapted to include the information that there were no observation events in each inter-event interval and that there were events at the observation times:
\[
\mathcal{L}(\mathbf{Q},\boldsymbol{\lambda};x_{[0,\tau]})=
\left[\prod_{t=1}^{T-1}
\exp\left\{(\mathbf{Q}-\boldsymbol{\Lambda})\Delta_t\right\}_{x_{\tau_t},x_{\tau_{t+1}}}\lambda_{x_{\tau_{t+1}}}
\right]
\exp\left\{(\mathbf{Q}-\boldsymbol{\Lambda})(\tau_e-\tau_T)\right\}_{x_{\tau_T},x_{\tau_{e}}}
\]
as the transition probability. Only imputing the states at observation times would reduce the amount of computation a little and still permits a Gibbs step for updating the $\mathbf{B}$ matrix.  As mentioned in the discussion, the evolution process occurs in continuous time and so it is natural to impute the full continuous-time process; however, there is also a more pragmatic reason to do this. If one only imputed the states at event times, one would be forced to perform inference using  Metropolis—Hastings for both $\mathbf{Q}$ and $\boldsymbol{\lambda}$,  and deal with the issues that we mention earlier in the supplementary material. The full underlying Markov chain is exactly what is needed for a Gibbs sampler on $\mathbf{Q}$ and $\boldsymbol{\lambda}$, and that is why our algorithm samples the full underlying Markov chain.

\section{Application: Additional results}

The results from the four-state Hidden Markov model appear in Table \ref{tab.fourstate}, and those for the three-state model but with  observation times no longer informative on the state are given in Table \ref{tab.threestatenotime}.

To calculate the marginal likelihood, integrating out the hidden Markov process, let $\mathbf{L}(t;\boldsymbol{\beta})$ be a diagonal matrix with $L_{k,k}=f(o_t|X_{\tau_t}=k)$. When the first event is guaranteed to occur at time $0$, the full likelihood for a the set of event times and observations for a single person is:
$$
\mathcal{L}(\boldsymbol{\beta},\boldsymbol{\lambda},\mathbf{Q},\boldsymbol{\nu})
\propto
\boldsymbol{\nu}^\top \mathbf{L}(0;\boldsymbol{\beta})
\exp[(\mathbf{Q}-\boldsymbol{\lambda})\tau_1]
\left\{\prod_{t=1}^T\boldsymbol{\Lambda}\mathbf{L}(t;\boldsymbol{\beta})\exp[(\mathbf{Q}-\boldsymbol{\Lambda})(\tau_{t+1}-\tau_t)]\right\}\mathbf{1},
$$
where $\tau_{T+1}=t_e$, the end of the observation window. For each of the three- and four-state CTMC-MMPP models we evaluate this expression at the model's posterior mean parameter estimate. For the BIC calculations, the three-state model has 14 parameters and the four-state model has 23; the number of patients is $n=1000$.

\begin{table}
	\caption{\label{tab.fourstate} Posterior median of four-state MMPP parameters associated with 95\% credible intervals.}
	\centering
	{\begin{tabular*}{40pc}{@{\hskip5pt}@{\extracolsep{\fill}}c@{}c@{}c@{}c@{\hskip5pt}}
			\hline
			Parameter  & State 1 &  State 2& State 3\\
			\hline
			$\lambda$ (Marked point process rate) &   4.82 (4.70, 4.96)& 5.51 (5.40, 5.62) & 7.71 (7.54, 7.89) \\
			$\nu$ (initial state distribution) & 0.33 (0.30, 0.37) & 0.35 (0.30, 0.40) & 0.32 (0.28, 0.36) \\
			Expected number of drugs dispensed (GP) & 0.70 (0.66, 0.75) & 4.25 (4.21, 4.28) & 8.66 (8.62, 8.70)\\
			Expected number of drugs dispensed (ED)& 0.43 (0.37, 0.50) & 4.01 (3.94, 4.08) & 8.65 (8.58, 8.73) \\
			Expected number of drugs dispensed (HOSP)&  0.51 (0.45, 0.58) &3.81 (3.73, 3.88) &   8.48 (8.39, 8.56)\\
			Expected number of drugs dispensed (SPEC)&  0.51 (0.45, 0.58) & 4.15 (4.07, 4.22)  & 8.65 (8.56, 8.74) \\
			$q_{1.}$ (Transition rate from State 1) &$-$&0.28 (0.24, 0.32)& 0.02 (0.01, 0.04) \\
			$q_{2.}$ (Transition rate from State 2)&$-$&$-$& 0.18 (0.17, 0.20)\\
			$q_{.4}$ (Transition rate to death state) &0.07 (0.06, 0.09) &0.07 (0.06, 0.08)&0.23 (0.21, 0.26)\\
			\hline
		\end{tabular*}
	}
\end{table}

\begin{table}
	\caption{\label{tab.threestatenotime} Posterior median of three-state CTHMM (with observation timing not informative on the state) parameters associated with 95\% credible intervals.}
	\centering
	{\begin{tabular*}{40pc}{@{\hskip5pt}@{\extracolsep{\fill}}c@{}c@{}c@{\hskip5pt}}
			\hline
			Parameter  & State 1 &  State 2\\
			\hline
			$\nu$ (initial state distribution)  & 0.76 (0.73, 0.79) & 0.24 (0.21, 0.27)  \\
			Expected number of drugs dispensed (GP) & 3.59 (3.55, 3.62) & 8.37 (8.33, 8.41) \\
			Expected number of drugs dispensed (ED)& 3.73 (3.65, 3.80) & 8.81 (8.73, 8.90)  \\
			Expected number of drugs dispensed (HOSP)&  3.32 (3.24, 3.40) &8.52 (8.43, 8.61) \\
			Expected number of drugs dispensed (SPEC)&  3.51 (3.43, 3.59) & 8.45 (8.35, 8.55)  \\
			$q_{1.}$ (Transition rate from State 1) &$-$&0.16 (0.14, 0.18) \\
			$q_{.3}$ (Transition rate to death state) &0.09 (0.08, 0.11) &0.04 (0.03, 0.05) \\
			
			\hline
		\end{tabular*}
	}
\end{table}

\bibliographystyle{chicago}
\bibliography{HMMPM}

\end{document}